\documentstyle[12pt]{article}
\title{Equilibrium transitions in finite populations of players} 
\author{Jacek Mi\c{e}kisz \\ Institute of Applied Mathematics \\
and Mechanics \\ Warsaw University  \\ ul. Banacha 2  \\ 02-097
Warsaw, Poland} 
\begin{document} 
\baselineskip=20pt
\maketitle 

\noindent {\bf Abstract}. 
We discuss stochastic dynamics of finite populations of individuals playing games. 
We review recent results concerning the dependence of the long-run behavior 
of such systems on the number of players and the noise level. In the case 
of two-player games with two symmetric Nash equilibria, when the number 
of players increases, the population undergoes multiple transitions between 
its equilibria.  
\vspace{3mm}

\vspace{5mm}

\eject

\section{Introduction}

\noindent Many socio-economic and biological processes can be modeled as systems 
of interacting individuals; see for example Santa Fe collection of papers 
on economic complex systems \cite{santa}, econophysics bulletin \cite{ekono}, 
and statistical mechanics and quantitative biology archives \cite{arch}. 

Here we will consider game-theoretic models of many interacting agents \cite{wei,hof,ams}. 
In such models, agents have at their disposal certain strategies and their payoffs 
in a game depend on strategies chosen both by them and by their opponents.
A configuration of a system, that is an assignment of strategies to agents, 
is a Nash equilibrium if for any agent, for fixed strategies of his opponents, 
changing the current strategy will not increase his payoff. 
One of the fundamental problems in game theory is the equilibrium selection
in games with multiple Nash equilibria. In two-player games with two strategies 
we may have two Nash equilibria: a payoff dominant (also called efficient) 
and a risk-dominant one. In the efficient equilibrium, players 
receive highest possible payoffs. The strategy is risk-dominant 
if it has a higher expected payoff against a player playing both strategies 
with equal probabilities. It is played by individuals averse to risks. 

One of the selection methods is to construct a dynamical system 
where in the long run only one equilibrium is played with a high frequency. 
Here we will discuss an adaptive dynamics introduced by Robson and Vega-Redondo
\cite{rvr}. In their model, at any time period, individuals play only one game 
with randomly chosen opponents (they do not play against an average strategy 
as in the replicator dynamics or the adaptive model of Kandori, Mailath, and Rob \cite{kmr}). 
The selection part of the dynamics ensures that if the mean payoff of a given strategy 
at the time $t$ is bigger than the mean payoff of the other one, then the number 
of individuals playing the given strategy should increase in $t+1$. 
In addition, with a small probability representing the noise of the system, 
players may make mistakes. 

To describe the long-run behavior of stochastic dynamics, Foster and Young \cite{foya} 
introduced a concept of stochastic stability. A state of a system 
(a number of individuals playing the first strategy in our models) is {\em stochastically stable} 
if it has a positive probability in the stationary state in the limit of zero noise. 
It means that in the long run we observe it with a positive frequency.

Here we review recent results concerning the dependence of the long-run behavior 
of the above desribed dynamics on the number of players and the noise level. 
We will combine these results to show that in the case of two-player games 
with two symmetric Nash equilibria, when the number of players increases, 
the population undergoes multiple transitions between its equilibria.  

\section{Adaptive dynamics with mistakes}

\noindent We will consider a finite population of $n$ individuals
who have at their disposal one of two strategies:
$A$ and $B$. At every discrete moment of time, $t=1,2,...,$
they are randomly paired (we assume that $n$ is even)
to play a two-player symmetric game with payoffs given by the following matrix:
\vspace{5mm}

\hspace{23mm} A \hspace{2mm} B   

\hspace{15mm} A \hspace{3mm} a \hspace{3mm} b 

U = \hspace{6mm} 

\hspace{15mm} B \hspace{3mm} c \hspace{3mm} d

where the $ij$ entry, $i,j = A, B$, is the payoff of the first (row) player when
he plays the strategy $i$ and the second (column) player plays the strategy $j$. 
We assume that both players are the same and hence payoffs of the column player are given 
by the matrix transposed to $U$; such games are called symmetric. 

An assignment of strategies to both players is a {\em Nash equilibrium}, if for each player, 
for a fixed strategy of his opponent, changing the current strategy will not increase his payoff. 
If $a>c$ and $d>b$, then $(A,A)$ and $(B,B)$ are two Nash equilibria. 
If $a+b<c+d$, then the strategy $B$ has a higher expected payoff against a player playing 
both strategies with equal probabilities. We say that $B$ risk dominates the strategy $A$ 
(the notion of the risk-dominance was introduced and thoroughly studied 
by Hars\'{a}nyi and Selten \cite{hs}). If in addition $a>d$, 
then we have a selection problem of choosing between the payoff-dominant 
(also caled efficient) equilibrium $(A,A)$ and the risk-dominant $(B,B)$.  

At every discrete moment of time $t$, the state of our population
is described by the number of individuals, $z_{t}$, playing $A$. 
Formally, by the state space we mean the set
$$\Omega=\{z, 0 \leq z\leq n\}.$$
Now we describe the dynamics of our system.
It consists of two components: selection and mutation.
The selection mechanism ensures that if the mean payoff 
of a given strategy, $\pi_{i}(z_{t}), i=A,B$, 
at the time $t$ is bigger than the mean payoff 
of the other one, then the number of individuals
playing the given strategy should increase in $t+1$. 

Let $p_{t}$ denote the random variable
which describes the number of cross-pairings, i.e. the number of pairs 
of matched individuals playing different strategies at the time $t$.
Let us notice that $p_{t}$ depends on $z_{t}$.
For a given realization of  $p_{t}$ and $z_{t}$,
mean payoffs obtained by each strategy are as follows:
\begin{equation}
\pi_{A}(z_{t},p_{t})=\frac{a(z_{t}-p_{t})+bp_{t}}{z_{t}},
\end{equation} 
$$\pi_{B}(z_{t},p_{t})=\frac{cp_{t}+d(n-z_{t}-p_{t})}{n-z_{t}},$$
provided $0<z_{t}<n$. 

We assume that at any time period, each individual 
has a revision opportunity with a small positive probability $\tau$ 
and adopts a strategy with the higher mean payoff. 
Players may make mistakes. At every time period, each player who has 
a revision opportunity, instead of following the selection rule may adopt the other strategy
with a small probability $\epsilon$. It is easy to see, 
that for any two states of the population, there is a positive probability
of the transition between them in some finite number of time steps. 
We have therefore obtained an irreducible Markov chain with $n+1$ states. 
It has a unique stationary state (a probability mass function) 
which we denote by $\mu^{\epsilon}_{n}.$ For any $z \in \Omega$,
$\mu^{\epsilon}_{n}(z)$ is the frequency of visiting the state $z$ in the long run.
The following definition was introduced by Foster and Young \cite{foya}.
\vspace{2mm}

\noindent {\bf Definition} $z \in \Omega$ is stochastically stable if 
$ \lim_{\epsilon \rightarrow 0}\mu_{n}^{\epsilon}(z)>0.$
\vspace{2mm}

\section{Equilibrium transitions}

\noindent We review here recent results concerning the dependence of stochastic stability 
of equilibria on the number of players.

They are based on a certain tree representation of stationary states of irreducible Markov chains
(\cite{freiwen1,freiwen2,shub}; see also Appendix). 
We assume that at any time period, each individual has a revision opportunity with a small
positive probability $\tau$. It follows that $z=0$ and $z=n$ are the only absorbing states. 
After a finite number of steps of the noise-free dynamics, we arrive 
at one of these two states and stay there forever - there are no other recurrence classes. 
Therefore to obtain a stationary state in the limit of zero noise, it is enough to count 
a number of mistakes the population needs to evolve between these states. 
If one requires, for example, fewer mistakes to evolve from $z=0$ to $z=n$ 
than from $z=n$ to $z=0$, then $z=n$ is stochastically stable. 

Robson and Vega-Redondo proved that for a sufficiently big number of players, 
the efficient strategy $A$ is stochastically stable \cite{rvr}.
They showed that $\lim_{\epsilon \rightarrow 0}\mu^{\epsilon}_{n}(n)=1$
which means that in the long run, in the limit of no mistakes, all individuals play $A$. 

However, their proof requires the number of players to be sufficiently big. 
It was showed in \cite{population} that the risk-dominant strategy $B$ is stochastically stable 
if the number of players is below $(2a-c-b)/(a-c)$.

Let us recall the proof. If the population consists of only one $B$-player and $n-1$ $A$-players 
and if $c>[a(n-2)+b]/(n-1)$, that is $n< (2a-c-b)/(a-c)$, 
then  $\pi_{B}> \pi_{A}.$ It means that one needs only one mistake 
to evolve from $z=n$ to $z=0.$ It is easy to see that two mistakes are necessary 
to evolve from $z=0$ to $z=n$ which finishes the proof.      

To see stochastically stable states, we need to take the limit of the zero noise level. 
It was showed in \cite{population} that for any arbitrarily low fixed noise level, 
if the number of players is big enough, then in the long run only 
a small fraction of the population plays the efficient strategy $A$. 
Smaller the noise level is, fewer individuals play $A$.  

Let us note that the above theorem concerns an ensemble of states, not an individual one. 
In the limit of the infinite number of players, that is the infinite number of states of the system, 
every single state has zero probability in the stationary state. It is an ensemble of states 
that might be stable. Ensemble and stochastic stability in spatial games with local interactions 
were recently discussed in \cite{statphys,physica,statmech}. For an interesting discussion 
on the importance of the order of taking different limits 
$(\tau  \rightarrow 0, n \rightarrow \infty,$ and $\epsilon \rightarrow 0)$
in evolutionary models (especially in the Aspiration and Imitation model) see \cite{samuel}. 

Now we combine the above theorems and obtain

\noindent {\bf Theorem}
\vspace{2mm}

For any $\delta >0$ and $\beta >0$ there exist $\epsilon(\delta, \beta)$
and $n_{1} < n_{2} < n_{3}(\epsilon) < n_{4}(\epsilon)$ such that 

if $n < n_{1}=\frac{2a-c-b}{a-c}$, then $\mu_{n}^{\epsilon}(z =0) > 1- \delta,$

if $n_{2} < n < n_{3}(\epsilon)$, then $\mu_{n}^{\epsilon}(z = n) > 1- \delta,$

if $n > n_{4}(\epsilon)$, then $\mu_{n}^{\epsilon}(z \leq \beta n) > 1- \delta$
for a sufficiently small $\tau$.

We see that for a fixed noise level, when the number of player increases, 
the population undergoes twice a transition between its two equilibria.
Of course, for any fixed number of players, $n>n_{2}$, if the noise level 
is sufficiently small, then almost all indivisuals will play 
in the long run the efficient strategy $A$. 

In order to study the long-run behavior of stochastic population dynamics,  
we should estimate the relevant parameters to be sure what limiting procedures 
are appropriate in specific examples. Equilibrium transitions in other 
stochastic dynamics of finite populations were recently investigated in 
\cite{nowak1,nowak2}. 
\vspace{4mm}

\noindent {\bf Appendix}
\vspace{2mm}

\noindent The following tree representation of stationary distributions 
of Markov chains was proposed by Freidlin and Wentzell \cite{freiwen1,freiwen2}, see also \cite{shub}.
Let $(\Omega,P)$ be an irreducible Markov chain with a state space 
$\Omega$ and transition probabilities given by $P^{\epsilon}: \Omega \times \Omega \rightarrow [0,1]$. 
It has a unique stationary distribution, $\mu^{\epsilon}$, also called a stationary state. 
For $X \in \Omega$, let an X-tree be a directed graph on $\Omega$ such that from every $Y \neq X$ 
there is a unique path to $X$ and there are no outcoming edges out of $X$. 
Denote by $T(X)$ the set of all X-trees and let 
\begin{equation}
q^{\epsilon}(X)=\sum_{d \in T(X)} \prod_{(Y,Y') \in d}P^{\epsilon}(Y,Y'),
\end{equation}
where the product is with respect to all edges of $d$. 
We have that
\begin{equation}
\mu^{\epsilon}(X)=\frac{q^{\epsilon}(X)}{\sum_{Y \in \Omega}q^{\epsilon}(Y)}
\end{equation}
for all $X \in \Omega.$

Let us assume now that after a finite number of steps of the noise-free dynamics, i.e. $\epsilon=0$, 
we arrive at one of two absorbing states, say $X$ and $Y$, and stay there forever - there are 
no other recurrence classes. It follows from the tree representation that any state different from absorbing states 
has zero probability in the stationary distribution in the zero-noise limit. 
Consider a dynamics in which $P^{\epsilon}(Z,W)$ for all $Z, W \in \Omega$, is of order $\epsilon^{m}$, 
where $m$ is the number of mistakes involved to pass from $Z$ to $W$ or is zero. Then one has to compute
the minimal number of mistakes, $m_{XY}$, needed to make a transition from the state $X$ 
to $Y$ and the number of mistakes, $m_{YX}$, to evolve from $Y$ to $X$. 
$q(X)$ is of order $\epsilon^{m(YX)}$ and $q(Y)$ is of order $\epsilon^{m(XY)}$.  
It follows that if $m_{XY} < m_{YX}$, then $Y$ is stochastically stable and moreover 
$ \lim_{\epsilon \rightarrow 0}\mu^{\epsilon}(Y)=1.$    


\vspace{2mm}

\begin{thebibliography}{99}

\bibitem{santa}{W. B. Arthur, S. N. Durlauf, and D. A. Lane, eds. 
{\em The Economy as an Evolving Complex System II}.
(Addison-Wesley, Reading, MA, 1997).}
\bibitem{ekono}{Econophysics bulletin on www.unifr.ch/econophysics}
\bibitem{arch}{Condensed Matter and Quantitative Biology archive on xxx.lanl.gov}

\bibitem{wei}{J. Weibull, {\em Evolutionary Game Theory} 
(MIT Press, Cambridge MA, 1995).}
\bibitem{hof}{J. Hofbauer and K. Sigmund, {\em Evolutionary Games and Population Dynamics} 
(Cambridge University Press, Cambridge, 1998).}
\bibitem{ams}{J. Hofbauer and K. Sigmund, Evolutionary game dynamics, 
{\em Bulletin AMS} {\bf 40} (2003), 479-519.} 

\bibitem{rvr}{A. Robson and F. Vega-Redondo, Efficient equilibrium selection
in evolutionary games with random matching, {\em J. Econ. Theory} {\bf 70} (1996), 65-92.}
\bibitem{kmr}{M. Kandori, G. J. Mailath, and R. Rob, 
Learning, mutation, and long-run equilibria in games,
{\em Econometrica} {\bf 61} (1993), 29-56.} 
\bibitem{foya}{D. Foster and P. H. Young,
Stochastic evolutionary game dynamics.
{\em Theoretical Population Biology} {\bf 38} (1990), 219-232.} 

\bibitem{hs}{J. Hars\'{a}nyi and R. Selten, 
{\em A General Theory of Equilibrium
Selection in Games} (MIT Press, Cambridge MA, 1988).} 

\bibitem{freiwen1}{M. Freidlin and A. Wentzell,
On small random perturbations of dynamical systems, {\em Russian Math. Surveys}
{\bf 25} (1970), 1-55.}
\bibitem{freiwen2}{M. Freidlin and A. Wentzell,
{\em Random Perturbations of Dynamical Systems} (Springer Verlag, New York, 1984).}
\bibitem{shub}{B. Shubert, A flow-graph formula for the stationary 
distribution of a Markov chain, {\em IEEE Trans. Systems Man. Cybernet.}
{\bf 5} (1975), 565-566.}

\bibitem{population}{J. Mi\c{e}kisz, Equilibrium selection in evolutionary games 
with random matching of players, {\em J. Theor. Biol.} {\bf 232} (2004), 23-32.} 

\bibitem{statphys}{J. Mi\c{e}kisz, Stochastic stability in spatial games, 
{\em J. Stat. Phys.} {\bf 117} (2004), 99-110.}
\bibitem{physica}{J. Mi\c{e}kisz, Stochastic stability in spatial three-player games, 
{\em Physica A} {\bf 343} (2004), 175-184}
\bibitem{statmech}{J. Mi\c{e}kisz, Statistical mechanics of spatial evolutionary games, 
{\em J. Phys. A: Math. Gen.} {\bf 37} (2004), 9891-9906.}

\bibitem{samuel}{L. Samuelson, {\em Evolutionary Games and Equilibrium Selection}.
(MIT Press, Cambridge MA, 1997).}

\bibitem{nowak1}{M. A. Nowak, A. Sasaki, C. Taylor, and D. Fudenberg, 
{\em Nature} {\bf 428} (2004), 646-650.}
\bibitem{nowak2}{C. Taylor, D. Fudenberg, A. Sasaki, and M. A. Nowak, 
Evolutionary game dynamics in finite populations, {\em Bull. Math. Biol.} {\bf 66} (2004), 1621-1644.}

\end{thebibliography}
\end{document}